\documentclass[showpacs,twocolumn,prl]{revtex4}

\usepackage{amsmath}
\usepackage{amssymb}
\usepackage[T1]{fontenc}
\usepackage{graphics}

\newcommand{\hb}{\bar{h}}

\newcommand{\vla}{\boldsymbol{\lambda}}

\newcommand{\kb}{k_{{\scriptscriptstyle B}}}

\newcommand{\Jt}{\widehat{J}}
\newcommand{\coul}{{\scriptscriptstyle{C}}}

\newcommand{\lm}{\lambda_\sigma}
\newcommand{\B}{\mathcal{B}}
\newcommand{\D}{\mathcal{D}}
\newcommand{\vk}{\mathbf{k}}
\newcommand{\vecr}{\mathbf{r}}
\newcommand{\vR}{\mathbf{R}}
\newcommand{\vh}{\mathbf{h}}
\newcommand{\vz}{\mathbf{0}}
\newcommand{\vm}{\mathbf{m}}

\newcommand{\ml}{\mathbf{\Lambda}}
\newcommand{\mlk}{\mathbf{\Lambda}(\vk; \vla)}
\newcommand{\mlmu}{\mathbf{\Lambda}^{-1}}
\newcommand{\Jst}{J_{\st}}
\newcommand{\Jsto}{J_{\st}^0}
\newcommand{\Jtst}{\widehat{J}_{\st}}
\newcommand{\Jtss}{\widehat{J}_{\sisi}}
\newcommand{\Gmn}{G_{\st}}
\newcommand{\Lx}{\Lambda_{{\scriptscriptstyle{X}}}}

\newcommand{\LN}{\Lambda_{\N}}
\newcommand{\LNk}{\LN (\vk; \vla)}
\newcommand{\LZ}{\Lambda_{\Z}}
\newcommand{\LZk}{\LZ (\vk; \vla)}

\newcommand{\ld}{\lambda^\dagger}
\newcommand{\ldc}{\lambda^\dagger_{c}}
\newcommand{\ldd}{\lambda^{\dagger^{{\scriptstyle{2}}}}}
\newcommand{\ldcd}{\lambda^{\dagger^{{\scriptstyle{2}}}}_c}

\newcommand{\dJ}{\Delta \bar{J}}
\newcommand{\dJdk}{\Delta J^\dagger(\vk)}
\newcommand{\Jtud}{\widehat{J}_{\ud}}
\newcommand{\dJud}{\Delta J_{\ud}}
\newcommand{\dJdd}{\Delta J_{\dd}}
\newcommand{\dJuu}{\Delta J_{\uu}}
\newcommand{\Jtudk}{\widehat{J}_{\ud}(\vk)}
\newcommand{\jz}{j_{0}}
\newcommand{\intk}{\int_{\vk}}

\newcommand{\Sd}{S_d}
\newcommand{\sisi}{\sigma  \sigma}
\newcommand{\st}{\sigma  \tau}
\newcommand{\ud}{+ -}
\newcommand{\uu}{+ +}
\newcommand{\dd}{- -}

\newcommand{\cOkd}{{\mathcal{O}}(k^2)}
\newcommand{\cOkq}{{\mathcal{O}}(k^4)}
\newcommand{\cO}{{\mathcal{O}}}
\newcommand{\sq}{\Sigma_4 (\uk)}

\newcommand{\Id}{{\mathcal{J}}_d}

\newcommand{\Tc}{T_{c}}
\newcommand{\roc}{\rho_c}
\newcommand{\N}{{\scriptscriptstyle N}}
\newcommand{\Z}{{\scriptscriptstyle Z}}
\newcommand{\NN}{{\scriptscriptstyle N N}}
\newcommand{\ZZ}{{\scriptscriptstyle Z Z}}
\newcommand{\NZ}{{\scriptscriptstyle N Z}}

\newcommand{\XY}{{\scriptscriptstyle X  Y}}
\newcommand{\SNN}{S_{\NN}}
\newcommand{\SZZ}{S_{\ZZ}}
\newcommand{\SNZ}{S_{\NZ}}
\newcommand{\SXY}{S_{{\scriptscriptstyle X Y}}}
\newcommand{\GNN}{G_{\NN}}
\newcommand{\GZZ}{G_{\ZZ}}
\newcommand{\GNZ}{G_{\NZ}}
\newcommand{\GXY}{G_{{\scriptscriptstyle X Y}}}
\newcommand{\xiN}{\xi_{\N}}
\newcommand{\xiNo}{\xi_{\N}^{\scriptscriptstyle{0}}}
\newcommand{\xiNu}{\xi_{{\N , {\scriptscriptstyle{1}}}}}

\newcommand{\xiNp}{\xi_{{\N , p}}} 
\newcommand{\xiNpdp}{\xi_{{\N , p}}^{2p}}
\newcommand{\xiZp}{\xi_{{\Z , p}}} 
\newcommand{\xiZpdp}{\xi_{{\Z , p}}^{2p}}

\newcommand{\xiNZpdp}{\xi_{{\NZ , p}}^{2p}}
\newcommand{\xiNi}{\xi_{{\N , \infty}}}

\newcommand{\xiZi}{\xi_{{\Z , \infty}}}
\newcommand{\xiZu}{\xi_{{\Z , {\scriptscriptstyle{1}}}}}

\newcommand{\xiZd}{\xi_{{\Z , {\scriptscriptstyle{2}}}}}
\newcommand{\xiD}{\xi_{{\scriptscriptstyle D}}}
\newcommand{\xiDc}{\xi_{{\scriptscriptstyle D} , c}}
\newcommand{\ns}{{\mathcal{S}}}
\newcommand{\fud}{\mbox{$\frac{1}{2}$}}

\newcommand{\fuq}{\mbox{$\frac{1}{4}$}}
\newcommand{\Ro}{R_{0}}
\newcommand{\Rod}{R_{0}^2}
\newcommand{\RN}{R_{\N}}

\newcommand{\RZ}{R_{\Z}}
\newcommand{\RZc}{R_{{\Z},c}}
\newcommand{\vir}{\, ,}
\newcommand{\pt}{\, .}

\newcommand{\Kbcc}{K_{bcc}}
\newcommand{\Io}{\mathcal{I}_{0}}
\newcommand{\IM}{\mathcal{I}^{max}}

\newcommand{\dJst}{\Delta J_{\st}}
\newcommand{\dJss}{\Delta J_{\sisi}}

\newcommand{\kl}{(\vk ; \vla)}
\newcommand{\Tr}{(T, \rho)}
\newcommand{\Trc}{(T_c, \rho_c)}

\newcommand{\uk}{\hat{\vk}}
\newcommand{\fic}{\varphi^{\coul}}
\newcommand{\dst}{\delta_{\sigma \tau}}
\newcommand{\das}{\delta_{{\scriptscriptstyle{J}}}}
\newcommand{\Trtc}{(T, \rho) \rightarrow (\Tc, \roc)}
\newcommand{\pims}{{{\textsc{BISM}}}}

\begin{document}

\title{Ionic criticality : an exactly soluble model}

\date{\today}

\author{Jean-No\"el Aqua}
\author{Michael E. Fisher}
\affiliation{Institute for Physical Science and Technology, University of 
Maryland, College Park, Maryland 20742, USA}

\begin{abstract}
Gas-liquid criticality in ionic fluids is studied in exactly soluble
spherical models that use interlaced sublattices to represent hard-core 
\textit{multi}component systems. Short range
attractions in the uncharged fluid drive criticality but charged 
ions do not alter the universality class. Debye screening remains exponential
\textit{at} criticality in charge-symmetric 1:1 models. However, 
\textit{asymmetry} couples 
charge and density fluctuations in a direct manner: the charge correlation 
length then diverges precisely as the density correlation length and 
the Stillinger-Lovett rule is violated \textit{at} criticality.  
\end{abstract}


\maketitle

The nature of gas-liquid (or, more generally fluid-fluid) criticality in 
systems in which long-range ionic interactions play a significant role has 
been a focus of attention since still-puzzling experiments questioned the 
appropriate universality class \cite{wein&schr01}. 
Beyond further experiments \cite{wein&schr01}, numerous theoretical
\cite{mef94,stel95,lee&mef96,mura00} 
and computational \cite{rome&orko00c} studies have been reported; however, 
basic questions remain open. Certainly, the character of criticality  
depends on the range of the interactions: One expects an Ising critical point 
in a fluid with short-range couplings but mean-field behavior for interactions
of sufficiently long-range. So, might the introduction of ions interacting via 
long-range Coulomb forces destroy an Ising-type critical point? Coulomb
interactions are exponentially screened in a conducting classical fluid, as 
proved rigorously at low densities \cite{Bryd&Fede80}. Charge fluctuations
in a fluid of $\ns$ species of charges $q_\sigma$ and valences 
$z_\sigma = q_\sigma / q$ (where $q$ is an elementary charge), thus decay 
over a few Debye screening lengths $\xiD$, where  
\begin{equation}
  \label{defxid}
  \xiD^{-2} (T, \rho) =   4 \pi  
  \rho q^2 \sum\nolimits_{\sigma=1}^{\ns} 
  z_\sigma^2 x_{\sigma} / \kb T  \vir
\end{equation}
in which $\rho = \sum_\sigma \rho_\sigma$ is the overall density while the 
mole fractions are $x_{\sigma} = \rho_\sigma / \rho$. But, does exponential 
screening on this scale hold near and at criticality?

Indeed, a major open issue is the behavior of the pairwize \textit{charge} 
correlations near the critical point, where the  \textit{density} fluctuations 
diverge strongly. With  
\begin{equation}
  \label{} \Gmn (\vecr)  =  \langle
  \rho_\sigma (\vz) \rho_\tau (\vecr) \rangle
  - \rho_\sigma \rho_\tau \vir
\end{equation}
the correlation functions, $\GNN$, $\GZZ$, and $\GNZ$ for the density, 
charge, and charge-density, are \cite{lee&mef96}
\begin{equation}
\label{} 
 \GXY (\vecr; T, \rho) = \sum\nolimits_{\sigma, \tau}
  q_\sigma^{\vartheta_{{\scriptscriptstyle X}}} 
  q_\tau^{\vartheta_{{\scriptscriptstyle Y}}}  \Gmn (\vecr; T, \rho)  \vir 
\end{equation}
where $X$ and $Y$ may be $N$ or $Z$ while $\vartheta_{\N} = 0$ and  
$\vartheta_{\Z} = 1$. Except at the critical point $\Trc$, itself,
we may suppose that the corresponding structure factor, $\SNN$, has the 
small $k = |\vk|$ expansion
\begin{equation}
  \label{dvlsnn}
  \SNN (\vk) / \SNN (\vz) = 1 + \sum\nolimits_{p\geq1} 
  (-)^{p} \xiNpdp \Tr \, k^{2p} \pt
\end{equation}
Near criticality, $\SNN (\vz; T, \rho_c)$ diverges as $1/t^\gamma$ 
when $t \equiv (T - \Tc)/\Tc \rightarrow 0+$, while the length 
$\xiNi$ characterizing the exponential decay of $\GNN (\vecr; T, \roc)$ 
diverges as $\xiNo / t^\nu$ (where short-range forces have been assumed). 
At criticality, density fluctuations are long-ranged and 
$\SNN (\vk; \Tc, \roc) \sim 1/k^{2-\eta}$. 

By contrast, the charge structure factor should obey 
\begin{equation}
  \label{dvlszz}
  \SZZ (\vk) = 0 + \xiZu^2  k^2 - 
  \sum\nolimits_{p\geq2} (-)^{p} \xiZpdp \Tr k^{2p} \vir
\end{equation}
where the first vanishing term results from electroneutrality 
reflecting the \textit{internal} screening in an ionic fluid, while  
the Stillinger-Lovett sum-rule \cite{stil&love68,mart88}
\begin{equation}
  \label{stillove}
  \xiZu \Tr = \xiD \Tr \vir
\end{equation}
characterizes the screening of \textit{external} charges. Does this hold near 
and at criticality? Finally, we focus also on the charge correlation length 
$\xiZi \Tr$, that specifies the exponential decay of $\GZZ (\vecr)$ when 
$r \rightarrow \infty$. How does $\xiZi$ vary when $\xiNi$ diverges near 
criticality?

\begin{figure}
\scalebox{0.21}{\includegraphics{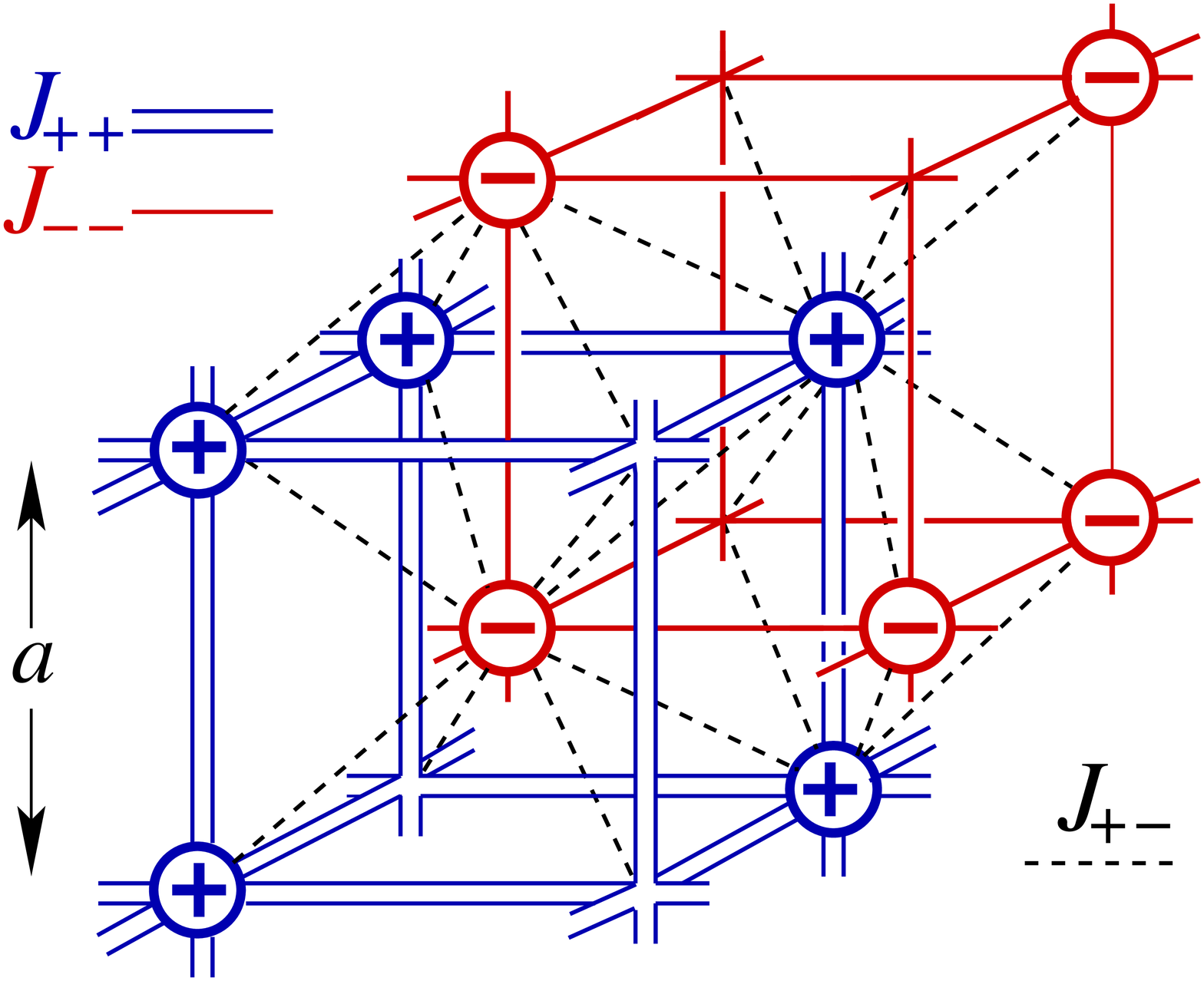}}
  \caption{\label{rezo3d} Interlaced $+$ and $-$ sc sublattices with 
    $a_0 = \fud \sqrt{3} a$.}
\end{figure}

To obtain insight into these questions, we study \textit{multicomponent} 
lattice gas generalizations of the spherical model 
\cite{joyc72,smit03,barb&mef91} specifically 
designed to represent hard-core interactions and thus
avoid the mutual ``annihilation'' of oppositely charged ions 
on the same site. This crucial feature, which (in contrast to 
\cite{smit03}) allows gas-liquid criticality to survive in the presence of 
Coulomb interactions, is accomplished by using a set of equivalent
interlacing sublattices (with sites $i$ at $\vR_i^\sigma$ with spacing $a$), 
one for each of the $\sigma = 1, 2, \ldots ,\ns$ distinct particle species 
\cite{cornu&janco87}. Thereby unlike charges cannot approach closer than an 
effective hard-core diameter $a_0$: see, e.g., Fig. \ref{rezo3d}.

To specify the models more fully, consider a multicomponent grand canonical 
lattice gas with site occupancy variables $n_i^\sigma = 0,1$; this is
equivalent to an Ising magnet with spins $s_i^\sigma=2 n_i^\sigma - 1 = \pm 1$,
subject to fields $h_\sigma$ (linearly related to the chemical potentials, 
$\mu_\sigma$). For attractive interparticle potentials, the corresponding 
spin-spin couplings, $J_{\st} (\vR)$, at lattice separations $\vR$, are 
positive. We decompose these couplings as 
\begin{equation}
  \label{}
  \Jst (\vR) = \Jsto (\vR) - \fuq q_\sigma q_\tau \fic (\vR) \vir
\end{equation}
where $\Jsto (\vR)$ represents short-range or, more generally, integrable 
interparticle interactions which we suppose suffice to drive gas-liquid 
criticality (even if 
the charges $q_\sigma$ vanish). We take the $d$-dimensional Coulomb potential, 
$\fic (\vR) \sim 1/R^{d-2}$ ($d>2$), as the solution of an appropriate 
discrete Laplace equation \cite{jn&mef03} (with, for convenience, a uniform 
neutralizing background so that electroneutrality, 
$\sum_\sigma x_\sigma q_\sigma = 0$, must be imposed). Fourier transforms, 
$\Jtst (\vk)$, are defined by summing over one sub-lattice, with 
Brillouin zone $\B$, and it is useful to introduce
\begin{equation}
  \label{}
  \dJst (\vk) \equiv \fud [ \Jtst (\vz) - \Jtst(\vk) ] \pt
\end{equation}

To model a simple 1:1 electrolyte one needs only $\ns = 2$ components, 
say $+$ and $-$ with $q_\pm = \pm q$. We may then identify basic energy and 
range scales, $\jz$ and $\Ro$, via 
\begin{equation}
  \label{defjz}
  \Jtud^{\, 0} (\vz) \equiv \kb T_0 \equiv 2 \jz >0 \vir
\end{equation}
and, assuming short-range isotropic nonionic couplings, 
\begin{equation}
  \label{}
  \dJuu^0 + 2 \dJud^0 + \dJdd^0 \approx 2 \jz \Ro^2 k^2 > 0  \vir
\end{equation}
as $k \rightarrow 0$. 
Now, if $J_{\uu} (\vR) = J_{\dd} (\vR)$, the model is (fully) 
\textit{charge symmetric}
(as is the well known continuum Restricted Primitive Model or RPM 
\cite{wein&schr01,mef94,stel95,rome&orko00c}). 
A suitable \textit{charge asymmetry} parameter is then 
\cite{jn&mef03}
\begin{equation}
  \label{defdas}
  \das = \max_{\vk \in \B} | \dJuu (\vk) - \dJdd (\vk) | / \kb T_0 \vir
\end{equation}
which might, e.g., be used to represent distinct ionic sizes, $a_+$ and 
$a_-$ 
[6(a)]. As the simplest ``\textit{Basic Ionic Model}''
it suffices to take only nearest neighbor couplings, $J_{\ud}^0 > 0$ and 
$J_{\uu}^0 = - J_{\dd}^0$; then one has $\jz = 2^{d-1} J_{\ud}^0$ and 
$\das = d |J_{\uu}^0| / 2^{d-2} |J_{\ud}^0|$. Finally, as a dimensionless
measure of the 
relative strength of the Coulomb interactions, it is helpful to introduce the 
\textit{ionicity} \cite{more&telo99}
\begin{equation}
  \label{defIo}
  \Io = q^2 / a^{d-2} \kb T_0 \pt
\end{equation}

Of course, even this Basic Ionic Model is insoluble for $d \geq 2$. 
Accordingly, in the standard way \cite{joyc72,smit03,barb&mef91}, we 
``sphericalize'' these multicomponent models by taking the spins $s_i^\sigma$
as unbounded continuous variables subject only to $\ns$ 
\textit{spherical constraints}, $\langle s_\sigma^2 \rangle = 1$, 
enforced with the aid of $\ns$ Lagrange multipliers which, for later 
convenience, we write as $\lambda_\sigma + \fud \Jtss (\vz)$. In full 
generality, the singular part of the total free energy 
$f[T, \vh; \vla(T, \vh)]$ in the thermodynamic limit, is then \cite{jn&mef03}
\begin{equation}
\label{} 
  f_s (T; \vla) = \fud \kb T \intk 
  \ln \left\{  \left| \mlk \right|  / (\kb T)^{\ns}  \right\} \vir
\end{equation}
where $\intk \equiv (a/2 \pi)^d \int_{\vk \in \B} d^d \vk$ and  
$\vh=(h_\sigma)$ and $\vla=(\lambda_\sigma)$, while the $\ns \! \times \! \ns$
matrix $\ml$ has elements 
\begin{equation} 
   \Lambda_{\st} =  \bigl[\lm + \dJss (\vk) \bigr] \dst
   - \fud (1- \dst) \Jt_{\sigma \tau} (\vk) \pt
\end{equation}
The field-dependent contribution to the free energy is 
\begin{equation}
\label{} 
  f_h  = - \fuq  \langle \vh | \mlmu (\vz; \vla) | \vh \rangle  \quad 
  \mbox{with} \quad \vh = 2 \ml (\vz; \vla) \vm \vir
\end{equation}
where $\vm=(m_\sigma)$ and 
$m_\sigma = \langle s_\sigma \rangle = 2 \langle n_\sigma \rangle - 1$
so that $\rho_\sigma = \fud \rho_\sigma^{max} ( 1 + m_\sigma)$. Finally, 
the Lagrange multipliers $\vla (T, \vh)$ are determined implicitly 
via the $\ns$ spherical constraints 
$\langle s_\sigma^2 \rangle = \left( \partial f / \partial \lm 
   \right) = 1$. 
These results are valid while the eigenvalues of $\ml$ remain positive;  
the vanishing of any one of them signals a thermodynamic singularity. 

For brevity hereon we focus on the two-species, 1:1 case (with 
$q_\pm = \pm q$), the $+$ and $-$ ions residing on one of two hypercubic 
sublattices displaced by $\fud (a, a, \cdots, a)$: see Fig.\ref{rezo3d} for 
$d=3$. The eigenvalues of $\ml$, to be called $\LZ$ and $\LN$ for reasons 
soon to be evident, are then simply the $+$ and $-$ roots of the quadratic
equation $| \ml - x \mathbf{I}| = 0$,  
\begin{gather}
  \label{defLx}
   \Lx (\vk; \vla) =  \bar{\lambda} + \dJ (\vk) 
   - (-)^{\vartheta_{{\scriptscriptstyle{X}}}} 
   \, \D (\vk; \vla) \quad \mbox{with} \\
   \D^2 (\vk; \vla) \equiv [ \ld + \dJdk ]^2 + \fuq \Jtud^{\, 2} (\vk) \vir
\end{gather}
where, for each variable $g_+$, $g_-$, $g_{\uu}$, etc., we have introduced the
mean $\bar{g} = \fud (g_+ + g_-)$ and difference 
$g^\dagger = \fud (g_+ - g_-)$, while the square root is chosen so that 
$\D (\vz; \vla) >0$ and $\Lx (\vk)$ is analytic.

A pivotal result now transpires \cite{jn&mef03}, namely, the linear 
decomposition of the structure factors via
\begin{equation}
\label{psxx}
  \frac{\SXY \kl}{\kb T /  4 \rho a^d} =
  \frac{B_{\XY}^{\N} (\vk; \vla) }{\LNk} 
  + \frac{B_{\XY}^{\Z} (\vk ; \vla)}{\LZk} \vir
\end{equation}
in which $B_{\NN}^{\N} = B_{\ZZ}^{\Z} = 1 - B_{\NN}^{\Z} = 1 - B_{\ZZ}^{\N}$, 
while
\begin{equation}
  B_{\NN}^{\N}  (\vk; \vla) =  \fud + \fuq | \Jtudk | / \D (\vk; \vla) \vir
\end{equation}
and $B_{\NZ}^{\N} + B_{\NZ}^{\Z} = 0$. In the \textit{charge symmetric cases}, 
all the $g^{\dagger}$ variables vanish by definition, so
 $B_{\NN}^{\N} \equiv 1$ and $B_{\NN}^{\Z} = B_{\ZZ}^{\N} \equiv 0$: this  
implies that $\SNN$ is entirely governed by $\LN$, and, likewise, $\SZZ$ by 
$\LZ$, so justifying the notation. Conversely, in \textit{non}symmetric models,
\textit{both} $\LN$ and $\LZ$ contribute to \textit{all} the structure 
functions. 

Finally, the spherical constraints reduce, first, to 
\begin{equation} 
  \label{csl}
  \kb T \Id (\vla) + \hb^2 / 4 \lambda^2 = 1   
  \quad \mbox{with} \quad \lambda = \bar{\lambda} -  \jz' \vir  
\end{equation}
where $\jz' = \fud \Jtud (\vz)$ \cite{jn&mef03}, while the basic integral 
\begin{equation}
  \label{defId}
  \Id (\vla) = \fuq \intk \left[ \LN^{-1} (\vk; \vla) + \LZ^{-1} (\vk; \vla)
  \right] \vir
\end{equation}
becomes singular, typically as $\Delta \Id \sim - \lambda^{1/\gamma}$, when 
$\lambda$ $(\sim t^\gamma)$ vanishes, where $\gamma$ is the appropriate 
critical exponent \cite{joyc72,barb&mef91,jn&mef03}; lastly 
$\ld$ is given (implicitly) by 
\begin{equation}
  \label{csld}
   \ld \intk 1/ \left|\ml (\vk; \vla)\right| = - \intk \dJdk / 
    \left|\ml (\vk; \vla)\right| \pt  
\end{equation}
Recall, however, that $\ld$ vanishes identically in charge symmetric cases:
more generally, this result relates $\ld$ to the asymmetry parameter $\das$ 
[see Eq. \eqref{defdas}] via 
\begin{equation}
  \label{zar}
  \ld_c \approx c_\delta [\Jst] \, \das \vir
\end{equation}
where, however, $c_\delta [\Jst]$ might vanish ``accidentally''.

We are now in a position to answer, with explanations, the questions posed 
after Eqns. \eqref{defxid}-\eqref{stillove}. For specific numerical results 
we will invoke the \textit{Basic Ionic Spherical Model} (\pims), i.e., the 
sphericalized version of the $d=3$, $\ns = 2$, 1:1 model with 
$\rho_+ = \rho_- = \fud \rho$ and nearest-neighbor interactions as set out 
with Eqns. \eqref{defjz}-\eqref{defIo}. When $\das = \Io = 0$ one readily 
finds from the vanishing of $\LN (\vz)$ when $\lambda \rightarrow 0$ 
\cite{joyc72,smit03,barb&mef91,jn&mef03} that standard spherical-model
criticality and scaling pertains, with exponents $\beta = \fud$ and 
$\gamma = 1 - \alpha = (2 - \eta)\nu$, where short-range (non-Coulomb) 
interactions \cite{footnote} lead to $\eta = 0$ and 
$\gamma = \max \{ 2/(d-2); 1 \}$ (for $d \geq 2$). 

For $\Io$ and $\das$ not too large, the \textit{same} situation 
prevails{\textemdash}contrary to speculations (for hard-core continuum ionic 
models) that lack of symmetry might lead to mean-field criticality 
\cite{stel95}. This follows most directly from the small $\vk$ behavior of 
the eigenvalues, namely, 
\begin{align}
  \LNk & = \lambda + \jz \RN^2 (\vla) \, k^2 + \cOkq \vir \nonumber \\
  \LZk & = (\Sd q^2 / 4 a^d) [ 1/k^2 + \RZ^2  + \cOkd ]  \vir \label{dvlL}
\end{align}
where $\Sd$ is the area of a unit $d$-sphere, while \cite{jn&mef03}
\begin{align}
\nonumber
   \RN^2 (\vla) & = \Rod -  \Sd' a^2 \Io - 2 a^2 \ldd / \Sd \Io \jz^2 \vir \\
   \RZ^2 (\vla, \uk) & = 
     2 a^2 (\lambda + \jz') / \Sd \Io \jz + a^2 \sq \label{defxim} \vir
\end{align}
in which $S_3'= \pi / 144$ and 
$\Sigma_4 \equiv \frac{1}{48} \sum_\alpha k_\alpha^4 / k^4$. The 
crucial feature, following directly from \eqref{defLx}, is that the Coulomb 
singularity, characterized by 
$\widehat{\varphi}^{\coul} (\vk) \, \sim \, 1 / k^2$, 
cancels out of $\LN$ exactly thanks to electroneutrality. (The absence of this 
possibility for $\ns = 1$ results in the destruction of gas-liquid criticality 
by Coulomb interactions 
[11(b)].)

To ensure the stability of the critical point, one also needs 
\cite{jn&mef03}: (i) $\Io < \IM_d$, with 
$\IM_3 = \frac{96}{11} \pi \simeq 2.77$; (ii) $\RN^2 > 0$, which restricts 
$\Io$ and $\das$ to the interior of an ellipsoid with a vertex at 
$\das = \Io = 0$, which, for $d=3$, is 
$\frac{1}{72} \das^2 + [ \frac{1}{72} \pi \Io - (\Rod/a^2) ]^2 < \Ro^4 / a^4$;
and (iv) the absence of competing minima in $\LN (\vk)$, which is 
satisfied for sufficiently small $\das$, specifically by 
$\das < 1 - \frac{1}{12} \pi \Io$ in the \pims. 
The solution $\ld (\lambda)$ of \eqref{csld} then varies 
smoothly when $\lambda \rightarrow 0$ and \eqref{zar} applies. By \eqref{csl}
criticality is restricted to $\bar{h}=0$ and occurs at 
$\kb \Tc = 1 / \Id (0, \ldc)$ and $\rho_c = a^{-d} = \fud \rho_{max}$. 
For the \pims, we find $\Tc \approx T_0 / \Kbcc$, to lowest order in $\das$ 
and $\Io$, with $\Kbcc \simeq 1.39$ \cite{joyc72}. It transpires 
\cite{jn&mef03} that $\Tc (\das)$ is a \textit{decreasing} function of the 
asymmetry in accord with recent simulations of hard-core continuum 
electrolyte models 
[6(a)] that, however, contradict various 
approximate theories. Furthermore, a term varying as $\Io^{3/2}$ (in $d=3$) 
appears \cite{jn&mef03}, in accord with \cite{more&telo99} and in analogy with 
\cite{kim&mef01}.

As regards the density correlation lengths, \eqref{psxx} yields 
\begin{equation}
  \label{}
  \xiNu \Tr = \RN \left[ \vla \Tr \right]
  \left[\jz / \lambda \Tr \right]^{1/2} \vir
\end{equation}
for all acceptable $\Io$ and $\das$; when $\rho = \rho_c$, this diverges 
as $\xiNo / t^\nu$. Furthermore, all higher moments [see \eqref{dvlsnn}], 
including the ``true'' correlation length $\xiNi$ \cite{mef&burf67}, 
satisfy $\xiNp / \xiNu \rightarrow 1$ when $\Trtc$. For the \pims, with 
small $\Io$ and $\das$ we have $\xiNo \approx a / \pi \Kbcc \simeq 0.229\, a$, 
close to the $d=3$, nearest neighbor Ising model value \cite{mef&burf67}.

By contrast, the near-critical \textit{charge correlation lengths} 
depend radically on symmetry. In \textit{charge symmetric models}, where 
$B_{\ZZ}^{\Z} = 1$, it follows from \eqref{psxx} and \eqref{dvlL} that the 
Stillinger-Lovett sum-rule \eqref{stillove} is valid for all fluid
regimes \textit{including} the critical point. However, the true charge 
screening length is given by 
\begin{equation}
  \label{}
  \xiZi \Tr = \RZ \Tr [ 1 + \cO (\Io^2) ] \vir
\end{equation}
where, from \eqref{defxim} we find $\RZ / \xiD \rightarrow 1$ as 
$\rho \rightarrow 0$ whereas near criticality one has 
$\RZc / \xiDc \approx 2 \sqrt{T_0 / \Tc}$; that yields 
$\RZc / \xiDc \simeq 2.36$ for the {\pims} (for $\Io$ and $\das$ not too 
large). Furthermore, when $t \rightarrow 0$, the screening length, 
$\xiZi (\rho_c, T)$ gains, in general, a singular correction factor 
$[1+c_{1-\alpha} t^{1-\alpha}]$ \cite{jn&mef03}. Up to $\cO(\Io)$, the higher 
moments of $\SZZ$ satisfy $(\xiZp)^p \approx \xiD \RZ^{p-1}$. 

On the other hand, in \textit{nonsymmetric cases} $B^{\N}_{\ZZ}$ in 
\eqref{psxx} does \textit{not} vanish : rather one has
\begin{equation}
  \label{}
  B_{\ZZ}^{\N} (\vk; \vla) = 
 4 \ldd k^4 a^4 / \Sd^2 \Io^2 \jz^2  [ 1 +\cOkd ] > 0  \pt
\end{equation}
Consequently, \textit{all} charge correlations become infected by the 
divergent density fluctuations controlled by $\LN (\vk)$. Nevertheless, 
because of the factor $k^4$, the Stillinger-Lovett relation
\eqref{stillove} remains valid in the fluid regime except \textit{at}
criticality where it fails and we find \cite{jn&mef03,footnote}
\begin{equation}
  \label{deflhdd}
  (\xiZu / \xiD)^2_c = 1 + w_c^2 \ldcd = 1 + w_c^2 c_\delta^2 \das^2 
  + \ldots\vir
\end{equation}
where $w^2 = 2 a^2 / \Sd \Io \RN^2 \jz^2$ and,  
recalling \eqref{zar}, we note that $c_\delta [\Jst] \neq 0$ for the 
\pims. This critical point failure implies a breakdown of full screening that 
is necessarily associated with slow decay of certain ionic correlations 
\cite{mart88}. 
Indeed, when $\Trtc$ the charge decay length $\xiZi$ \textit{asymptotically
approaches} the density correlation length, $\xiNi$, and thus \textit{diverges}
as $\xiNo/t^\nu$ (for $\rho = \rho_c$). 

However, the fourth charge 
correlation moment is given (except \textit{at} criticality) by 
\begin{equation}
  \label{axizd}
  \xiZd^4 \Tr = \xiD^2  [\RZ^2 - \xiNu^2 w^2  \ldd]  \vir
\end{equation}
so that $|\xiZd (\rho_c, T)|$ diverges more weakly as $1/t^{\nu/2}$. Note also 
that on approaching criticality, $\xiZd^4$ changes sign (with respect to the 
symmetric case); for small $\das$ this crossover occurs in the {\pims} at 
$t = t_\times \approx 0.265 \, \das$. More generally the higher 
moments in nonsymmetric systems satisfy 
$ \xiZpdp \approx -  \xiNi^{2(p-1)}  \xiDc^2 w_c^2 \ldd_c$  leading to a 
hierarchy 
of critical exponents $|\xiZp| \sim 1 / t^{\nu (1-1/p)}$. Notwithstanding the
divergence of the charge correlation length, $\xiZi \approx \xiN 
\rightarrow \infty$, the charge-charge \textit{pair correlation function}, 
$\GZZ (\vecr)$, decays \textit{exponentially at} $\Trc$! Indeed on 
approach to criticality we obtain 
\begin{equation}
  \label{}
  \GZZ (\vecr) \varpropto  2 \das^2 \, \frac{T_0}{\Tc} \,  
  \frac{c_\delta^2}{\jz^2} \, \frac{\xiD^4}{\RN^2} \, 
  \frac{e^{-r/\xiN}}{\xiN^4 \Tr r} -  
  \frac{\xiD^2}{\RZ^4} \, \frac{e^{-r/\RZ}}{r} \vir
\end{equation}
to leading orders for $d=3$. At $\Tc$ 
only the second term survives since the first vanishes as $t^{4 \nu}$. 

Finally, as regards the cross-correlations embodied in the charge-density 
structure factor $\SNZ$, we find $\SNZ(\vk\!=\!\vz; T, \rho) \equiv 0$
except at $(\Tc, \rho_c)$ where the value $(\ldd \xiD^2 / \jz \RN^2)_c$ is 
realized. Moreover, on defining moments in analogy to \eqref{dvlszz}, one 
obtains
\begin{equation}
  \label{}
  \xiNZpdp \Tr \approx ( \ld \xiD^2 / \jz \RN^2 ) \xiNi^{2p} \Tr \vir
\end{equation}
near criticality; of course, all these moments vanish identically in charge 
symmetric models since $\ld \equiv 0$. 

\begin{table}
\caption{\label{resume} Charge correlation lengths near criticality 
where $\xiN$ diverges, while $\RZ/\xiD=\cO(1)$. The ionicity, 
$\Io \varpropto q^2$, and asymmetry factor $w_c c_\delta \das$, 
are defined in \eqref{defIo}, \eqref{zar} and  \eqref{deflhdd}.}
\begin{ruledtabular}
\begin{tabular}{lll}
  & $ $ \quad \quad  charge symmetric & $ $ 
  \quad \quad non-symmetric\\ 
 $\xiZu$ &$=\xiD=(4 \pi \rho q^2/\kb  T)^{1/2},$& 
 $=\xi{{\scriptscriptstyle D}}$ \, 
 for $\Tr \neq \Trc$,\\
  & $ $ \quad \quad \,  the Debye length, & $> \xiDc$ at $\Trc$,\\  
     $\xiZd^4$ &$=\xiD^2 \RZ^2 \Tr = \cO(\xiD^4),$ 
     &$=-\xiD^2[(w_c c_\delta \das)^2\xiN^2-\RZ^2]$,\\ 
  $\xiZp^{2p}$ &$=\xiD^2\RZ^{2(p-1)}[1+\cO(\Io)]$,& 
  $\approx -(w_c c_\delta \das)^2\xiD^2\xiN^{2(p-1)}$,\\ 
  $\xiZi$ &$=\RZ[1\!+\!\cO (\Io^2)]= \cO (\xiD)$,& 
  $\approx \xiN \sim 1/t^\nu$. 
\end{tabular}
\end{ruledtabular}
\end{table}

In summary, we have analyzed a class of exactly soluble spherical models for 
1:1 ionic systems and shown that the Coulomb interactions do not change the 
gas-liquid critical universality class{\textemdash}contrary to some 
suggestions \cite{wein&schr01,stel95}. The couplings between the charge 
correlations and the divergent critical density fluctuations 
\cite{jn&mef03,footnote} follow mainly from a remarkable structure-function 
decomposition, Eq. \eqref{psxx}, that respects the Stillinger-Lovett (SL) sum
rule (unlike \cite{mura00}). Our principal results are collected in Table 
\ref{resume}~: they are broadly consistent with Ornstein-Zernike-based 
arguments advanced for hard-core continuum electrolytes \cite{stel95}. In 
\textit{charge-symmetric} models density fluctuations are not directly coupled
to two-point charge correlations which remain of short-range and obey SL near 
and \textit{at} criticality; but in more realistic \textit{nonsymmetric} 
systems the density fluctuations ``infect'' the charge correlations which 
hence exhibit the same diverging correlation length. Moreover, the SL rule is 
then violated \textit{at} criticality \cite{jn&mef03,footnote} indicating an 
anomalous conducting state \cite{mart88}.

\begin{acknowledgments}
 The authors are grateful to F. Cornu, B. Jancovici, E.R. Smith and G. Stell 
 for 
 correspondence, and to Y. C. Kim for his interest. Support from the 
 \textsc{NSF} (under grants CHE 99-81772 and 03-01101) and assistance to 
 J.-N.A. from the French Ministry of Foreign Affairs under the 
 Lavoisier Fellowship program, is gratefully acknowledged. 
\end{acknowledgments}


\begin{thebibliography}{22}
\bibitem{footnote}{Significant differences that arise when 
$J_{\sigma \tau }^0(\mathbf {R}) \sim 1/R^{d+\varsigma }$, implying $\eta>0$ 
when $\varsigma<2$, will be described elsewhere.}
\end{thebibliography}
\end{document}